# Vagus nerve stimulation as a modulator of feedforward and feedback neural transmission

**Shinichi Kumagai[1,2], Tomoyo Isoguchi Shiramatsu[2], Kensuke Kawai[1], Hirokazu Takahashi[2*]**

[1]Department of Neurosurgery, Jichi Medical University, Tochigi, Japan; [2]Department of Mechano-Informatics, Graduate School of Information Science and Technology, The University of Tokyo, Tokyo, Japan

**\*Correspondence:**

Hirokazu Takahashi, Ph.D., Department of Mechano-Informatics, Graduate School of Information Science and Technology, The University of Tokyo, 7-3-1 Hongo, Bunkyo, Tokyo 113-8656, Japan. Tel: +81-3-5841-0461. Email: takahashi@i.u-tokyo.ac.jp

**Abstract**

Vagus nerve stimulation (VNS) has emerged as a promising therapeutic intervention across various neurological and psychiatric conditions, including epilepsy, depression, and stroke rehabilitation; however, its mechanisms of action on neural circuits remain incompletely understood. Here, we present a novel theoretical framework based on predictive coding that conceptualizes VNS effects through differential modulation of feedforward and feedback neural circuits. Based on recent evidence, we propose that VNS shifts the balance between feedforward and feedback processing through multiple neuromodulatory systems, resulting in enhanced feedforward signal transmission. This framework integrates anatomical pathways, receptor distributions, and physiological responses to explain the influence of the VNS on neural dynamics across different spatial and temporal scales. VNS may facilitate neural plasticity and adaptive behavior through acetylcholine and noradrenaline (norepinephrine), which differentially modulate feedforward and feedback signaling. This mechanistic understanding serves as a basis for interpreting the cognitive and therapeutic outcomes across different clinical conditions. Our perspective provides a unified theoretical framework for understanding circuit-specific VNS effects and suggests new directions for investigating their therapeutic mechanisms.

**Introduction**

Vagus nerve stimulation (VNS) has emerged as a promising therapeutic approach for various neurological and psychiatric disorders. This neuromodulatory technique, which involves electrical stimulation of the vagus nerve, has gained significant attention due to its broad therapeutic potential. The U.S. Food and Drug Administration (FDA) has approved VNS for several therapeutic applications. Its primary indication is treatment of drug-resistant epilepsy. The efficacy of VNS in reducing seizure frequency and severity has been demonstrated in numerous clinical trials in patients with intractable epilepsy (George et al., 1995) (Handforth et al., 1998) (Englot et al., 2011). In addition to epilepsy, VNS has also received FDA approval for the management of treatment-resistant depression. Studies have shown promising results in alleviating depressive symptoms in patients who do not respond adequately to conventional therapies including medication and psychotherapy (Rush et al., 2005) (Schlaepfer et al., 2008) (Aaronson et al., 2013) (Aaronson et al., 2017). Non-invasive VNS has been approved for the acute treatment of migraine, and clinical trials have demonstrated pain relief after attacks with stimulation (Tassorelli et al., 2018) (Martelletti et al., 2018). More recently, VNS has been approved as an adjunct therapy in stroke rehabilitation. VNS, when paired with motor rehabilitation, can enhance motor function recovery in stroke survivors (Dawson et al., 2016) (Dawson et al., 2021) (Kwakkel and Dobkin, 2021). These clinical effects are likely mediated through the modulation of various neurotransmitters including acetylcholine, noradrenaline, and serotonin (Krahl et al., 1998) (Raedt et al., 2011) (Fornai et al., 2011) (Furmaga et al., 2011) (Zhang et al., 2019) (Hulsey et al., 2016) (Engineer et al., 2019) (Hulsey et al., 2019) (Collins et al., 2021).

The potential benefits of VNS extend beyond the approved indications. Numerous studies have suggested that VNS may positively affect cognitive function. Studies have explored its impact on various aspects of cognition, including memory, executive function, and attention, in patients with epilepsy and mild cognitive impairment, and in healthy adults (Mertens et al., 2022) (Aniwattanapong et al., 2022) (Wang et al., 2022) (Klaming et al., 2022). However, it is crucial to acknowledge that the effects of VNS on cognitive processes are complex and not uniformly positive. While many studies have reported improvements, some have found no significant changes or even potential impairments in certain cognitive domains following VNS treatment (Clark et al., 1999) (Helmstaedter et al., 2001) (Ghacibeh et al., 2006) (McGlone et al., 2008) (Vonck et al., 2014) (Mertens et al., 2020) (Kong et al., 2024). These varied outcomes underscore the complex interactions between VNS and cognitive processes.

This perspective aims to explore the intricate relationship between VNS and brain information processing. We propose a novel framework to elucidate the mechanisms underlying the diverse effects of VNS on the neural circuitry.

**A novel framework for circuit-specific VNS effects**

Although numerous studies have demonstrated the efficacy of VNS in various neurological and psychiatric conditions, the underlying neural processes are not fully understood. Our recent studies provided initial insights into the layer-specific effects of VNS on sensory processing. We demonstrated that VNS predominantly enhanced auditory-evoked responses in the superficial layers of the primary auditory cortex (A1), with diminishing effects in the deeper layers (Takahashi et al., 2020) (Figure 1A-D). Furthermore, we found that VNS modulates oscillatory activities in A1 through the cholinergic and noradrenergic systems, suggesting distinct modulation of cortical oscillations (Kumagai et al., 2023) (Figure 1E). Additionally, our preliminary data showing changes in functional connectivity between the core and belt regions in the auditory cortex provides further evidence for the pathway-specific effects of VNS (Figure 1F).

Specifically, VNS enhanced the transfer entropy of evoked activity from the core to the belt regions, suggesting a strengthening of feedforward information flow in the auditory cortex. Based on these findings and several lines of supporting evidence, we propose a novel hypothesis: VNS modulates the balance between feedforward and feedback processing, specifically enhancing feedforward information flow in the thalamo-cortical and cortico-cortical systems.

This modulation can be interpreted within the framework of predictive coding, which proposes that the brain continuously generates predictions about incoming sensory inputs and minimizes the difference (prediction error) between these predictions and the actual inputs (Heilbron and Chait, 2018). Predictive coding has been mathematically formalized under the free energy principle, which quantifies the discrepancy between the brain's internal models and actual sensory inputs using variational Bayesian inference. According to this theory, prediction errors propagate from lower to higher areas in feedforward pathways, whereas predictions propagate in the opposite direction in feedback pathways (Friston and Kiebel, 2009) (Bastos et al., 2012) (Aitchison and Lengyel, 2017). In the auditory system, each level of processing compares incoming acoustic signals with what the brain expects to hear, generating prediction errors when there are mismatches between expected and actual sounds. These prediction errors flow upward from the primary auditory cortex through the higher auditory regions, whereas expectations about upcoming sounds flow downward. For example, when we listen to speech and hear an unexpected pronunciation or an unfamiliar accent, the brain initially generates large prediction errors; as we continue to listen, these prediction errors trigger an update of our predictions about the speaker's speech patterns, gradually reducing subsequent prediction errors and improving our ability to understand the speaker's speech.

In this predictive coding framework, VNS may enhance the feedforward signaling of prediction errors with respect to feedback signaling of prediction, thereby influencing how effectively the brain processes sensory information and updates its internal models. Multiple lines of evidence support this hypothesis at anatomical, physiological, and functional levels.

**Anatomical basis for circuit-specific modulation**

To understand how VNS might differentially affect feedforward and feedback processing, it is essential to consider the underlying anatomical organization of the cortical circuits. The cerebral cortex exhibits a hierarchical organization defined by distinct laminar patterns of inter-areal connections that support predictive coding processes (Figure 2A). In the cortical hierarchy, feedforward projections originate primarily from the supragranular layers (layers 2/3) and target layer 4, while feedback projections arise from the infragranular layers (layers 5/6) and terminate in layers 1 and 6 (Bastos et al., 2012) (Felleman and Van Essen, 1991) (Shipp, 2007). Although detailed anatomical studies have shown that both supragranular and infragranular layers contain feedforward and feedback streams, feedforward and feedback pathways predominantly use the supragranular and infragranular layers, respectively (Markov et al., 2014).

The supragranular layers exhibit precise point-to-point connectivity, enabling accurate signal transmission between specific cortical regions, which is crucial for feedforward sensory processing (Markov et al., 2013). In contrast, the infragranular layers show more diffuse connectivity patterns, facilitating broad signal integration and divergence across multiple areas, which is particularly important for feedback signaling (Markov et al., 2014). This dual organization allows the cortex to simultaneously achieve both precise and rapid information processing through targeted feedforward pathways, and flexible contextual interpretation through distributed feedback networks. Such anatomical organization provides the structural basis for predictive coding, where prediction errors are primarily conveyed through supragranular layers from lower to higher areas, whereas predictions are transmitted through infragranular layers from higher to lower areas (Friston and Kiebel, 2009).

VNS predominantly enhances neural responses in superficial layers (I-IV), with diminishing effects in deeper layers (Takahashi et al., 2020). This laminar specificity of VNS effects, combined with the anatomical organization described above, provides mechanistic evidence that VNS selectively enhances feedforward information flow in the cortical circuits.

**Receptor distribution of neuromodulatory transmitters**

The proposed mechanism can be further supported by examining the detailed anatomical laminar organization of the neuromodulatory systems. The distribution of cholinergic and noradrenergic receptors shows distinct layer-specific patterns, which aligns with this hypothesis.

The cholinergic inputs show a denser distribution in superficial layers (I-IV) associated with feedforward processing, with both nicotinic and M1 muscarinic receptors enriched in these layers (Palomero-Gallagher and Zilles, 2019). This organization suggests a potential mechanism by which VNS selectively modulates feedforward pathways.

The adrenergic receptor distribution exhibits intricate layer-specific patterns that may contribute to the proposed feedforward enhancement. $\alpha1$- and $\alpha2$-adrenergic receptors tend to be densely distributed in the superficial layers across many cortical regions, although their specific distribution patterns vary between different cortical areas (Palomero-Gallagher and Zilles, 2019). β-adrenergic receptors are predominantly localized in the deep cortical layers (V and VI) of the sensory cortex of young cats (Liu et al., 1993). During development, this distribution pattern shifts, ultimately establishing an adult pattern with two distinct bands of high receptor density in superficial and deep cortical layers.

The varying affinities of different adrenergic receptor subtypes (α2 highest, followed by α1, and then β) suggest that VNS may produce concentration-dependent effects. Indeed, noradrenaline exhibits dual effects on glutamate-evoked neuronal discharge through distinct receptor subtypes: α1-receptor activation results in facilitation, whereas β-receptor activation leads to the suppression of these glutamate-evoked responses (Devilbiss and Waterhouse, 2000). Moderate noradrenaline concentrations may preferentially activate α1-receptors in superficial layers, enhancing feedforward signaling, whereas higher concentrations could additionally recruit β-receptors in deep layers, potentially suppressing the influence of feedback pathways. Consistent with the suppressive role of the β-receptor, a pharmacological study has shown that β-receptor activation impairs prefrontal cortical cognitive function in both rats and monkeys (Ramos et al., 2005).

VNS-induced cortical plasticity exhibits an inverted-U relationship with stimulation intensity, where moderate intensities optimize plasticity, whereas both low and high intensities are less effective (Morrison et al., 2019). This relationship may be explained by the differential activation of the adrenergic receptor subtypes. α- and β-adrenergic receptors can exert opposing effects on synaptic plasticity in a concentration-dependent manner (Salgado et al., 2012) (Hays et al., 2023), suggesting that balanced activation of these receptor systems is crucial for optimal VNS-induced plasticity. Indeed, α2-receptor activation in the motor cortex is necessary for VNS-induced plasticity at a moderate (0.8mA) stimulation intensity (Tseng et al., 2021). Collectively, these findings suggest that VNS influences cortical processing through multiple receptor-specific mechanisms that depend on both the stimulation intensity and anatomical distribution of receptor subtypes.

**VNS-induced modulation of neural oscillations**

Supporting our hypothesis that VNS differentially modulates feedforward and feedback processing, electrophysiological evidence has demonstrated specific alterations in neural oscillations and information transmission patterns. Neural oscillations exhibit distinct characteristics across different spatial scales and laminar distributions. Gamma oscillations tend to coordinate local circuit activities such as sensory processing and are predominantly generated in the superficial cortical layers. In contrast, slower oscillations, such as theta waves, are more prominent in the deeper layers and facilitate communication between distant brain regions (Bastos et al., 2012) (Sarnthein et al., 1998) (Miller et al., 2018). These oscillations can interact hierarchically through cross-frequency coupling, providing a mechanism for integrating bottom-up sensory information with top-down cognitive control (Fries, 2023).

VNS enhances auditory-evoked gamma power through the cholinergic system and decreases theta power through the noradrenergic system in the rat auditory cortex (Kumagai et al., 2023). These data suggest that VNS strengthens feedforward pathways through the cholinergic system, while attenuating feedback circuits via the noradrenergic system. The VNS-induced enhancement of gamma oscillations indicates increased feedforward signaling, which is consistent with established evidence that gamma band activity mediates feedforward information flow (Bastos et al., 2015) (Chao et al., 2018) (Bastos et al., 2020) (Vezoli et al., 2021). However, the functional significance of VNS-induced theta suppression requires further investigation. Although theta rhythms may contribute to both feedforward and feedback processing (Bastos et al., 2015) (Chao et al., 2018) (Bastos et al., 2020) (Vezoli et al., 2021), they are commonly linked to top-down control (Sarnthein et al., 1998) (Cavanagh and Frank, 2014), which involves the coordination of distributed neural networks to implement cognitive functions such as working memory (Sarnthein et al., 1998). In this context, VNS-induced enhancement of gamma oscillations coupled with theta suppression may represent a shift in cortical processing that favors feedforward sensory transmission while reducing top-down feedback influences, potentially optimizing the balance between bottom-up and top-down information flow.

Further evidence comes from an EEG study of non-invasive VNS in healthy subjects, which decreased theta and alpha power while increasing beta and gamma power, indicating a shift toward cortical activation (Lewine et al., 2019). Consistent with these findings, intraoperative electrocorticographic recordings in patients with refractory epilepsy demonstrated that VNS significantly enhanced high-frequency spectral power, particularly in the beta and gamma bands (Yokoyama et al., 2020). A recent multicenter study examining intracranial recordings in patients with epilepsy who had both VNS and responsive neurostimulation (RNS) systems demonstrated significant reductions in the theta-band power during VNS (Ernst et al., 2023). Similarly, transcutaneous VNS leads to pupil dilation and attenuation of occipital alpha oscillations, suggesting increased arousal and potentially enhanced sensory processing (Sharon et al., 2021).

However, a replication study observed increased pupil size but failed to replicate the alpha attenuation (Lloyd et al., 2023), and an early study of invasive VNS in epilepsy patients found no significant effects on awake EEG background rhythms (Salinsky and Burchiel, 1993). Furthermore, both invasive and non-invasive VNS modulate low-frequency spectral power across distributed cortical networks, but the effects vary considerably based on stimulation parameters and individual differences (Schuerman et al., 2021).

While these mixed findings of oscillatory activity suggest variable effects on resting-state brain activity, more consistent effects emerge when examining active sensory processing, likely due to VNS modulation of engaged feedforward and feedback circuits in response to sensory stimuli. Information theoretical analyses have demonstrated that VNS rapidly enhances the feature selectivity and information transmission of thalamic neurons in the rat somatosensory system (Rodenkirch and Wang, 2020). This improvement coincided with the suppression of the thalamic burst firing. Given that the thalamus plays a crucial role in relaying sensory information to the cortex, these findings suggest that VNS specifically enhances feedforward sensory processing through more precise and efficient feedforward signaling from the thalamus to the primary sensory cortex.

Together, these electrophysiological findings strongly support our hypothesis that VNS differentially modulates feedforward and feedback neural transmission, primarily through enhancement of feedforward pathways. This modulation appears to occur through rapid activation of distinct neuromodulatory systems, potentially explaining the observed effects on sensory processing.

**Neuromodulatory control of neural plasticity**

The understanding of neuromodulatory control over cortical plasticity has evolved significantly over the past three decades. Early investigations in the 1990s demonstrated that sensory experience alone was insufficient to drive cortical plasticity and revealed that both sensory experience and neuromodulatory signaling are essential to induce long-lasting changes in cortical circuits (Bakin and Weinberger, 1996) (Kilgard and Merzenich, 1998a) (Kilgard and Merzenich, 1998b). These pioneering insights into neuromodulatory mechanisms formed the basis for subsequent investigations into VNS-induced plasticity. Our hypothesis of VNS-enhanced feedforward processing also builds upon these foundational studies of neuromodulatory control over cortical plasticity: enhancement of the feedforward pathway, i.e., prediction error, would promote plasticity and update generative models in higher-order cortical areas.

Early studies have demonstrated that the cholinergic system plays a crucial role in cortical plasticity, finding that pairing auditory stimuli with nucleus basalis stimulation induces long-lasting changes in auditory cortical receptive fields (Bakin and Weinberger, 1996). This form of plasticity shared key features with behavioral memory: it was associative, specific, and long-lasting (McLin et al., 2002) (Weinberger, 2003) (Weinberger, 2004) (Weinberger, 2007). Nucleus basalis stimulation paired with tones induces large-scale reorganization of frequency representation in the auditory cortex (Kilgard and Merzenich, 1998a) (Kilgard and Merzenich, 1998b), triggering a sequence of synaptic changes: rapid disinhibition followed by delayed enhancement of excitation (Froemke et al., 2007). Paring sensory stimuli with nucleus basalis stimulation enhances sensory processing in two ways: it improves the reliability of neuronal responses and reduces correlations between cortical neurons, resulting in improved perceptual performance in behaving animals (Goard and Dan, 2009) (Froemke et al., 2013). Similarly, VNS paired with tones induced targeted plasticity in the auditory cortex and eliminated behavioral correlates of tinnitus in noise-exposed rats (Engineer et al., 2017) (Engineer et al., 2011).

Like cholinergic modulation, noradrenergic signaling from the locus coeruleus (LC) powerfully influences cortical plasticity through distinct mechanisms. The activation of LC neurons can trigger long-lasting changes in auditory cortical responses (Martins and Froemke, 2015). Specifically, pairing tones with LC stimulation induced coordinated plasticity in both the modulatory and sensory pathways, leading to improved auditory perception that could last for weeks. Pairing auditory stimuli with LC stimulation induced two distinct patterns of plasticity in auditory cortical neurons: frequency-selective changes with an increase or decrease in evoked responses in more than 30% of neurons, and non-selective response changes across frequencies in approximately 50% of neurons (Edeline et al., 2011). This suggests that LC activation can simultaneously drive both stimulus-specific refinement and broader changes in the cortical excitability. In a more recent study, LC activation was found to facilitate cochlear implant-driven plasticity, significantly enhancing long-term perceptual performance in a rat model of cochlear implantation (Glennon et al., 2023). Initially, LC activation induced a widespread increase in auditory cortical responses, enhancing the activity to both paired and non-rewarded inputs. Over the course of training, these changes became progressively more selective, with LC activity increasingly favoring paired stimuli while suppressing responses to non-rewarded inputs.

While cholinergic modulation suppresses stimulus-evoked inhibition followed by selective enhancement of excitation to paired stimuli (Froemke et al., 2007) (Froemke et al., 2013), noradrenergic modulation enhances the overall gain of cortical responses, showing increased activity to both paired and unpaired stimuli, with some preference for the paired frequency (Martins and Froemke, 2015). These distinct modulatory effects suggest different roles in cortical plasticity: selective modification of specific inputs by acetylcholine versus broader changes in cortical processing by noradrenaline (Martins and Froemke, 2015) (Froemke, 2015). These studies on cholinergic and noradrenergic modulation help explain how VNS promotes plasticity through the engagement of these neuromodulatory systems (Hulsey et al., 2016) (Hulsey et al., 2019) (Hays et al., 2023) (Tseng et al., 2021) (Brougher et al., 2021) (Bowles et al., 2022) (Martin et al., 2024). These effects likely occur through vagal activation, first by engaging noradrenergic pathways, which then modulate cholinergic function, rather than through direct vagal-cholinergic projections. (Collins et al., 2021).

These plasticity effects induced by VNS (or cholinergic activation) are dependent on the timing of the paired movement or stimulus (Bowles et al., 2022) (Martin et al., 2024). As synaptic plasticity, i.e., potentiation or depotentiation in response to incoming input, should be effectively driven by prediction errors (Saponati and Vinck, 2023), the timing-dependent nature of the VNS effects suggests that cholinergic modulation emphasizes prediction error signals. The effectiveness of plasticity also depends on spatial and temporal precision and salience relative to the background neural activity, specifically during behaviorally relevant moments. Cholinergic signaling is likely to modulate normalization, through which neuronal activity is divided by the pooled activity of surrounding neurons, which is associated with reductions in noise correlations, minimization of trial-to-trial variability, and enhancement of the signal-to-noise ratio (Carandini and Heeger, 2011) (Schmitz and Duncan, 2018). Through these normalization effects, cholinergic modulation potentially refines the neural representation of prediction error signals, allowing neural circuits to update their predictive models more efficiently, based on experience.

**Dual neuromodulatory systems in behavioral strategy control**

The selective engagement of cholinergic and noradrenergic systems by VNS has important implications for adaptive behavior. Through its concurrent effects on these neuromodulatory systems, VNS can dynamically adjust the balance between feedforward sensory processing and feedback predictions based on contextual demands. In novel environments, VNS may optimize learning through two complementary mechanisms: cholinergic enhancement of feedforward prediction errors promotes effective model updating, whereas noradrenergic modulation facilitates adaptive model revision through adjusted state transition precision (Shine, 2019) (Munn et al., 2021) (Shine et al., 2021). This dual modulation benefits situations that require new learning or adaptation. However, in stable environments where established internal models guide behavior, VNS-induced enhancement of feedforward processing might unnecessarily increase the sensitivity to prediction errors, potentially disrupting performance.

These modulatory effects also influence the exploration-exploitation trade-off in behavior; exploitation facilitates focused task performance, while exploration promotes sampling of alternative behaviors (Aston-Jones and Cohen, 2005). The noradrenergic system operates in two distinct modes that regulate behavioral strategies (Aston-Jones and Cohen, 2005). In the phasic mode, neurons exhibit transient activation patterns time-locked to task-relevant decision processes, which promote exploitation by facilitating focused attention and optimal task performance when reward utility is high. In contrast, the tonic mode is characterized by a sustained elevation of baseline activity, which emerges when reward utility diminishes. This elevated tonic activity enhances responsiveness to task-irrelevant stimuli, thereby promoting the exploration of alternative behavioral opportunities. These distinct firing patterns may differentially engage in neural processing within the cortical hierarchy. Moderate tonic LC activation preferentially engages higher-order associative regions, whereas burst-like stimulation enhances the activity in lower-order sensory regions (Grimm et al., 2024). These differential effects on cortical processing may contribute to flexible adjustments in behavioral strategies, although precise mapping between experimental stimulation patterns and naturally occurring LC activity patterns requires further investigation. Indeed, VNS parameters can differentially modulate LC firing modes. While standard VNS (10-30 Hz) induces consistent activation of specific LC neurons, bursting VNS (300-350 Hz) promotes synchronous firing between LC neurons, although detailed analysis of temporal firing patterns in individual LC neurons remains to be elucidated (Farrand et al., 2023).

The cholinergic system enables efficient task performance by modulating sensory gain in task-relevant neural circuits through precise spatiotemporal control over cortical information processing. In the basal forebrain, two distinct populations of cholinergic neurons differentially contribute to behavioral control through different firing patterns. Burst-firing cholinergic neurons generate precisely timed responses to behaviorally salient events and can synchronously activate cortical circuits, while regular-firing neurons show distinct temporal activity patterns that correlate with successful performance in sensory detection tasks (Laszlovszky et al., 2020). Recent studies have shown that VNS may engage these cholinergic mechanisms by robustly modulating basal forebrain activity with precise temporal dynamics (Bowles et al., 2022) (Martin et al., 2024). Specifically, VNS-induced activation of basal forebrain neurons begins during stimulation and can persist for several seconds, leading to the activation of cortically projecting cholinergic axons and subsequent behavioral improvements (Martin et al., 2024).

Based on these distinct neuromodulatory mechanisms, VNS may dynamically regulate behavioral strategies through parallel engagement of both systems: noradrenergic control determines behavioral state transitions, while cholinergic modulation enhances task-specific neural processing. These dual mechanisms can be conceptualized within the energy landscapes of cortical activity, a probabilistic framework representing the stability of brain states, with noradrenergic effects flattening the landscape to enable flexible transitions, whereas cholinergic modulation deepens specific valleys to stabilize newly established brain states (Munn et al., 2021). This mechanistic framework suggests that optimal VNS application requires careful consideration of both the task demands and environmental stability. Although VNS might enhance learning and adaptation in novel environments, its application during periods requiring stable performance could be counterproductive.

**Cognitive implications of VNS-induced circuit modulation**

In a recent meta-analysis, invasive VNS produced limited cognitive benefits in patients with epilepsy, showing no significant improvements in overall cognitive performance, executive function, attention, and memory (Kong et al., 2024). Similarly, transcutaneous auricular VNS (taVNS) demonstrated only a small effect size (Hedges' $g$ = 0.21, 95% CI = .12-.29) on overall cognitive performance in healthy individuals (Ridgewell et al., 2021). These heterogeneous effects of VNS on cognitive function might be understood through differential modulation of feedforward and feedback processing. While VNS consistently enhances feedforward signals through cholinergic activation, this enhancement does not uniformly translate to improved cognitive performance, possibly due to the fundamental importance of balanced feedforward and feedback signals in cognitive processing (Shine et al., 2021) (Halvagal and Zenke, 2023).

The cognitive impact of VNS appears to be highly dependent on task context. In well-practiced tasks such as recalling memorized information or performing familiar motor skills, the brain predominantly utilizes established internal models through feedback processing. Under these conditions, VNS-enhanced feedforward signaling may introduce unnecessary interference by increasing the sensitivity to prediction errors. This interference has been observed in verbal memory tasks and cognitive flexibility assessments that rely on established internal representations (Ghacibeh et al., 2006) (Mertens et al., 2020). Conversely, VNS may enhance performance during new skill acquisition or environmental adaptation such as learning a new language or mastering novel motor patterns, as demonstrated in a study in which taVNS significantly improved adults' ability to learn novel letter-sound correspondences in unfamiliar orthographies (Thakkar et al., 2020). VNS enhances reinforcement learning, particularly in individuals with lower extraversion traits (Weber et al., 2021). The mechanism may involve modulation of the balance from internally generated representations toward externally driven sensory inputs, i.e., from prediction to prediction error.

The variable effects of VNS on cognitive function may stem not only from the task context but also from the stimulation parameters used across studies (Vonck et al., 2014). Cognitive benefits may be achieved at lower current intensities than those typically employed in epilepsy and depression (Broncel et al., 2020). Studies using implanted VNS systems have demonstrated that moderate stimulation intensities around 0.4-0.5 mA have been shown to enhance recognition memory, while higher intensities (0.75-2.5 mA) either fail to improve or may actually impair memory function (Clark et al., 1999) (Helmstaedter et al., 2001). This inverted U-shaped response pattern was observed in VNS-induced hippocampal plasticity. VNS optimally facilitates long-term potentiation (LTP), a form of synaptic plasticity underlying learning and memory, in the dentate gyrus of rats at 0.4 mA, while both lower (0.2 mA) and higher (0.8 mA) stimulation intensities produce weaker effects (Zuo et al., 2007). This intensity-dependent pattern appears to be mediated by differential levels of noradrenaline release, suggesting that VNS parameters effective for seizure control may inhibit feedback processing through excessive noradrenergic modulation.

The temporal precision of the VNS application has emerged as a critical factor in determining its cognitive effects. Unlike animal studies, many human studies have applied VNS without pairing with a specific task-relevant stimulus (Kong et al., 2024). VNS efficacy may be optimized when precisely synchronized with specific cognitive operations (Bowles et al., 2022) (Martin et al., 2024). This precise timing is crucial for optimal acetylcholine release, which is necessary to enhance task-relevant feedforward signaling that updates the internal model. VNS, as well as stimulation of the cholinergic nucleus basalis, paired with a specific frequency tone, induces cortical map expansion in a region corresponding to the paired tone frequency in the auditory cortex (Kilgard and Merzenich, 1998a) (Engineer et al., 2011), which improves perceptual learning but is not necessary for improved tone discrimination (Reed et al., 2011). A similar cortical map expansion is observed at the early stage of learning in auditory operant conditioning (Takahashi et al., 2010) (Takahashi et al., 2011), and this map expansion is associated with the diversity of tone-evoked neural responses (Takahashi et al., 2013). Early learning primarily relies on bottom-up sensory pathways (Makino and Komiyama, 2015); therefore, it may benefit from acetylcholine-mediated enhancement induced by VNS (Bowles et al., 2022) (Martin et al., 2024). Late-phase learning, which requires the formation of novel neural activity patterns through feedback processing, may not benefit from VNS enhancement of feedforward circuits (Carroll et al., 2024). Thus, the temporal precision of VNS with respect to task-relevant stimuli may be critical at the early stage of learning.

In contrast to timing-dependent effects, VNS may also fundamentally enhance perceptual precision through the modulation of arousal systems, independent of specific task demands (Collins et al., 2021). Both the noradrenergic and cholinergic systems activated by VNS induce widespread cortical excitation and strongly influence the arousal state. Heightened arousal levels enhance both single-neuron and population-level sensory encoding, characterized by increased signal-to-noise ratios and reduced noise correlations, while altering cortical activity patterns by reducing low-frequency oscillations and enhancing gamma-band synchronization (Vinck et al., 2015). Different levels of arousal also trigger distinct patterns of cholinergic modulation. During moderate arousal, acetylcholine release becomes more coordinated across cortical regions, whereas high-arousal states lead to profound decorrelation of cholinergic signals (Lohani et al., 2022). Such state-dependent changes in cholinergic signaling may optimize cognitive processing by dynamically adjusting the cortical network coordination across various behavioral states.

These insights suggest that optimizing VNS for cognitive enhancement requires careful consideration of multiple factors, including cognitive context, stimulation parameters, and temporal dynamics. Future applications should focus on precisely targeting VNS to specific phases of cognitive operations where enhanced feedforward processing would be most beneficial, while avoiding periods where it might disrupt established processing patterns. This approach may help to resolve the current heterogeneity in cognitive outcomes and lead to more effective therapeutic applications of VNS.

**Clinical implications of VNS-induced circuit modulation**

The enhancement of feedforward signaling by VNS may facilitate the update of internal models and provide a mechanistic framework for understanding its diverse therapeutic applications (Figure 2B). While this feedforward enhancement may be particularly relevant for conditions requiring internal model updating such as depression and stroke rehabilitation, VNS could also modulate aberrant feedback signaling that contributes to seizure propagation in epilepsy. This balanced modulation of the feedforward and feedback pathways helps restore optimal circuit dynamics across different neurological and psychiatric conditions.

**Epilepsy**

VNS treatment outcomes in patients with epilepsy remain notably heterogeneous, with the responder rate (≥50% reduction in seizure frequency) reaching approximately 60% within 2-3 years of therapy (Kawai et al., 2017). Although VNS has been in clinical use for several decades, reliable predictive markers of therapeutic response remain elusive (Workewych et al., 2020) (Clifford et al., 2024). Recent evidence suggests that alterations in thalamocortical connectivity may play a crucial role in determining the treatment outcomes. A diffusion tensor imaging (DTI) study in 56 children showed that VNS responders demonstrated greater preoperative integrity of thalamocortical pathways than non-responders (Mithani et al., 2019). A functional magnetic resonance imaging (fMRI) study in 21 medically intractable pediatric epilepsy patients reported that presurgically enhanced thalamocortical connectivity has also been associated with favorable VNS responses (Ibrahim et al., 2017). Indeed, an early PET imaging study demonstrated that increased thalamic blood flow during acute VNS was strongly correlated with improved seizure outcomes, supporting the thalamus as a key therapeutic target (Henry et al., 1999). The relationship between thalamocortical connectivity and VNS response provides compelling evidence for a circuit-based therapeutic mechanism. Further studies are needed to elucidate how VNS specifically alters information flow between the thalamus and cortex in patients with epilepsy.

Beyond thalamocortical circuits, recent laminar recordings during human seizures have revealed distinct patterns of cortical layer engagement during seizure initiation and propagation (Bourdillon et al., 2024). In the seizure onset zone, epileptic activity originates and persists in the infragranular layers, suggesting that the pathological activity patterns in these deep layers may drive seizure initiation. In areas outside the seizure onset zone, seizure activity predominantly engages layer I before descending to deeper layers, which corresponds to the anatomical pathway of the cortical feedback signals. Within this framework, VNS may exert its therapeutic effect by enhancing activity in the supragranular layers, which could help to counterbalance the pathological activity in the infragranular layers. The delayed therapeutic effect often observed with VNS might reflect the time required for the gradual modulation of these laminar-specific circuit dynamics through chronic stimulation.

**Depression**

Major depressive disorder (MDD) is a complex disorder involving multiple pathophysiological alterations, including changes in brain structure and function, inflammation, gut-brain axis, and hypothalamic-pituitary-adrenal axis, with at least 30% of patients showing an inadequate response to multiple medications (Marx et al., 2023) (McIntyre et al., 2023). While the monoamine hypothesis has historically dominated our understanding of MDD focusing on serotonin and noradrenaline systems, this framework cannot fully account for the delay between pharmacological action and clinical improvement, or the heterogeneous treatment responses. Although VNS exerts anxiolytic and antidepressant-like behavioral effects through serotonergic and noradrenergic pathways in rats (Furmaga et al., 2011), the underlying mechanism of action in MDD remains unclear (Nemeroff et al., 2006) (Grimonprez et al., 2015) (Carreno and Frazer, 2017).

fMRI studies have demonstrated that MDD involves disrupted network regulation across multiple brain regions, rather than dysfunction in any single neurotransmitter system. MDD is associated with hyperconnectivity within the default mode network, which is involved in self-referential processes and maladaptive depressive rumination (Hamilton et al., 2011) (Kaiser et al., 2015) (Hamilton et al., 2015) (Kawakami et al., 2024). Moreover, MDD exhibits a distinctive pattern of network imbalance: decreased connectivity between the frontoparietal network (involved in cognitive control) and dorsal attention network (involved in external attention), along with increased connectivity between the frontoparietal network and default mode network (involved in internal thought processes) (Kaiser et al., 2015). This suggests that the core features of depression may not simply arise from increased or decreased activity in specific regions but rather from an imbalance in network interactions. Such network dysregulation may help explain two characteristic features of depression: an excessive focus on internal thoughts and reduced engagement with the external environment.

Supporting this view, a resting-state fMRI study revealed that patients with MDD show reduced functional connectivity within both the visual and auditory networks, as well as between these networks, indicating altered sensory processing (Lu et al., 2020). Based on these network-level disruptions in MDD, VNS may exert its therapeutic effects by restoring the balance between internal and external processing. Specifically, by enhancing feedforward information flow through neural circuits, VNS could help shift attention from internal rumination toward external environmental inputs. Indeed, research has demonstrated that taVNS can significantly modulate default mode network connectivity in patients with MDD, suggesting its potential role in normalizing the disrupted network interactions characteristic of depression (Fang et al., 2016).

**Stroke rehabilitation**

Beyond its application in psychiatric disorders, VNS has emerged as a promising therapeutic approach in neurological rehabilitation. Clinical trials have demonstrated the efficacy of VNS paired with rehabilitation for improving upper limb function in patients with upper limb impairment at least 9 months after ischemic stroke (Dawson et al., 2021) (Kimberley et al., 2018). A pivotal randomized, triple-blind trial showed that VNS paired with rehabilitation significantly improved arm function compared to rehabilitation with sham stimulation, with 47% of VNS-treated patients achieving clinically meaningful responses in the Fugl-Meyer Assessment-Upper Extremity (FMA-UE) score versus 24% in the control group (Dawson et al., 2021). The therapeutic protocol involves precise timing of VNS lasting 0.5s with specific rehabilitation movements including reach and grasp exercises. This improvement in chronic stroke patients, with a mean time since stroke of approximately 3 years, is particularly noteworthy because functional improvement at this chronic stage is unexpected and this time point corresponds to the onset of the greatest functional decline in stroke survivors (Dhamoon et al., 2009). Recent long-term follow-up data also demonstrate that VNS-induced improvements not only persist, but continue to increase over the years (Francisco et al., 2023).

Animal studies have elucidated the neurobiological mechanisms underlying these clinical benefits. VNS drives task-specific plasticity in the motor cortex with paired VNS movement training, leading to a significant expansion of movement representations in motor maps (Porter et al., 2012). In a rat model of stroke, VNS enhances synaptic connectivity in motor networks and doubles motor recovery compared with rehabilitation alone (Meyers et al., 2018). This VNS-dependent plasticity appears to involve cholinergic signaling, as lesions of the nucleus basalis, the primary source of cortical acetylcholine, prevent VNS-dependent enhancement of motor cortex plasticity, and cortical cholinergic depletion blocks VNS-driven motor and sensory recovery (Hulsey et al., 2016) (Meyers et al., 2019). Furthermore, VNS promotes skilled motor learning via cholinergic reinforcement, mediated by selective modulation of outcome-relevant neural circuits in the motor cortex (Bowles et al., 2022). Its therapeutic benefit is optimized when stimulation is precisely timed with successful movements (Khodaparast et al., 2014). Through cholinergic modulation, VNS appears to enhance sensorimotor feedforward signaling, which enables more efficient updating of internal models, potentially explaining its ability to drive targeted plasticity and improve functional recovery in stroke rehabilitation. This mechanism may be particularly effective in chronic stroke, in which maladaptive feedback processing could interfere with motor learning and recovery.

In stroke rehabilitation, VNS is likely to support the reorganization of the somatopic map in the sensorimotor cortex. Such cortical reorganization may be a common mechanism for the treatment of other pathological conditions in the cortex. For example, deafferentation due to hearing loss disrupts the tonotopic map in the auditory cortex, resulting in tinnitus (Mühlnickel et al., 1998) (Eggermont, 2015) (Wake et al., 2024), and similarly the disorganization of somatopic maps in the somatosensory cortex is associated with phantom pain following limb amputation (Flor et al., 1995). VNS-assisted reorganization of these pathological cortices may be a potential treatment in the future (Engineer et al., 2011) (De Ridder et al., 2021).

**Methodological considerations in evaluating VNS interventions: comparing icVNS and taVNS**

The clinical evidence discussed above for epilepsy, depression, and stroke rehabilitation has primarily been established using invasive cervical VNS (icVNS). When evaluating VNS interventions for these diseases, it is crucial to distinguish between icVNS and non-invasive taVNS. While icVNS directly stimulates vagal nerve fibers, taVNS delivers stimulation through the skin of the external ear, potentially affecting multiple neural pathways beyond the vagus nerve (Mahadi et al., 2019). This anatomical and mechanistic distinction has important implications for the therapeutic efficacy. A recent commentary has highlighted that combining these distinct interventions in meta-analyses may lead to misleading conclusions regarding their relative effectiveness, emphasizing the need for intervention-specific evaluation of clinical outcomes (Malakouti et al., 2024). Both interventions activate the nucleus tractus solitarius (NTS), which receives the majority of vagal afferents but with notably different patterns across its subdivisions. The dorsolateral regions show stronger activation with icVNS, whereas the dorsomedial regions respond more robustly to taVNS (Ali et al., 2024). Detailed electrophysiological investigations have further demonstrated that, while taVNS and icVNS produce comparable overall NTS activation patterns, their effects at the single-neuron level can be markedly different and sometimes opposite (Owens et al., 2024). Of particular interest is the finding that taVNS produces more pronounced activation of the spinal trigeminal nucleus (Sp5), a structure known to receive direct projections from the auricular branch of the vagus nerve, than NTS activation. These observations provide strong evidence that these interventions may engage different neuronal pathways to achieve therapeutic effects. Intriguingly, transcutaneous stimulation of the cervical region has been shown to deactivate Sp5 (Frangos and Komisaruk, 2017), suggesting potentially different mechanisms of action between the cervical and auricular approaches. This modulatory pattern may help explain the therapeutic benefits of non-invasive cervical VNS in acute migraine attacks (Goadsby et al., 2017) (Brennan and Pietrobon, 2018) (Ashina, 2020) (Puledda et al., 2023), although the precise mechanisms remain to be fully elucidated.

These findings underscore a critical point: while icVNS and taVNS may appear to activate similar brain regions at the macro level such as fMRI (Badran et al., 2018), their effects on specific brainstem nuclei follow distinct patterns. This nuclei-specific targeting indicates the need for careful consideration when interpreting the therapeutic mechanisms of non-invasive VNS approaches, and highlights that we should not assume complete mechanistic equivalence with invasive VNS, even though they partially share some clinical benefits.

**Conclusion**

The hypothesis of the VNS-induced differential modulation of neural connections offers a novel perspective on the mechanisms underlying this neuromodulatory intervention. By selectively amplifying the feedforward signaling of prediction errors while weakening the influence of feedback circuits, VNS might facilitate the brain's ability to minimize prediction errors and optimize its internal model. This mechanism is consistent with the organization of neuromodulatory systems in the cerebral cortex, particularly cholinergic and noradrenergic projections, which are predominantly affected by VNS. This integrative view provides a potential explanation for the diverse effects of VNS and opportunities for computational research on neuromodulation. Future investigations focusing on the layer-specific actions of VNS, its interactions with neuromodulatory systems, and its impact on predictive brain processes may lead to more targeted and effective therapies. Taken together, this perspective underscores the potential of VNS as both a therapeutic tool and an approach for investigating the fundamental principles of neural information processing.

**Author contributions**

Shinichi Kumagai: Investigation, Writing – original draft, and Visualization. Tomoyo Isoguchi Shiramatsu: Methodology and Funding acquisition. Kensuke Kawai: Supervision and Funding acquisition. Hirokazu Takahashi: Conceptualization, Supervision, Writing – review & editing, and Funding acquisition.

**Funding**

This work was supported by JSPS KAKENHI (23H03465, 23H04336, 23H03023, 23KJ1866, 24H01544), AMED (24wm0625401h0001), JST (JPMJPR22S8), the Asahi Glass

Foundation, and the Secom Science and Technology Foundation.

**Conflict of Interests**

The authors declare that there are no conflicts of interest associated with this manuscript.

**Acknowledgements**

We thank Rie Hitsuyu, Kenji Ibayashi, Kotaro Ishizu, and Akane Matsumura for performing the animal experiments related to the concepts discussed in this perspective. In preparing this manuscript, Claude 3.5 Sonnet by Anthropic was used to improve readability. After using this tool, the authors reviewed and edited the content.

**References**

Aaronson, S.T., Carpenter, L.L., Conway, C.R., Reimherr, F.W., Lisanby, S.H., Schwartz, T.L., et al. (2013). Vagus nerve stimulation therapy randomized to different amounts of electrical charge for treatment-resistant depression: acute and chronic effects. *Brain Stimul* 6, 631-640. doi: 10.1016/j.brs.2012.09.013

Aaronson, S.T., Sears, P., Ruvuna, F., Bunker, M., Conway, C.R., Dougherty, D.D., et al. (2017). A 5-Year Observational Study of Patients With Treatment-Resistant Depression Treated With Vagus Nerve Stimulation or Treatment as Usual: Comparison of Response, Remission, and Suicidality. *American Journal of Psychiatry* 174, 640-648. doi: 10.1176/appi.ajp.2017.16010034

Aitchison, L., and Lengyel, M. (2017). With or without you: predictive coding and Bayesian inference in the brain. *Current Opinion in Neurobiology* 46, 219-227. doi: 10.1016/j.conb.2017.08.010

Ali, M.S.S., Parastooei, G., Raman, S., Mack, J., Kim, Y.S., and Chung, M.K. (2024). Genetic labeling of the nucleus of tractus solitarius neurons associated with electrical stimulation of the cervical or auricular vagus nerve in mice. *Brain Stimul* 17, 987-1000. doi: 10.1016/j.brs.2024.08.007

Aniwattanapong, D., List, J.J., Ramakrishnan, N., Bhatti, G.S., and Jorge, R. (2022). Effect of Vagus Nerve Stimulation on Attention and Working Memory in Neuropsychiatric Disorders: A Systematic Review. *Neuromodulation: Technology at the Neural Interface* 25, 343-355. doi: 10.1016/j.neurom.2021.11.009

Ashina, M. (2020). Migraine. *New England Journal of Medicine* 383, 1866-1876. doi: doi:10.1056/NEJMra1915327

Aston-Jones, G., and Cohen, J.D. (2005). An integrative theory of locus coeruleus-norepinephrine function: adaptive gain and optimal performance. *Annu Rev Neurosci* 28, 403-450. doi: 10.1146/annurev.neuro.28.061604.135709

Badran, B.W., Dowdle, L.T., Mithoefer, O.J., LaBate, N.T., Coatsworth, J., Brown, J.C., et al. (2018). Neurophysiologic effects of transcutaneous auricular vagus nerve stimulation (taVNS) via electrical stimulation of the tragus: A concurrent taVNS/fMRI study and review. *Brain Stimulation* 11, 492-500. doi: 10.1016/j.brs.2017.12.009

Bakin, J.S., and Weinberger, N.M. (1996). Induction of a physiological memory in the cerebral cortex by stimulation of the nucleus basalis. *Proc Natl Acad Sci U S A* 93, 11219-11224. doi: 10.1073/pnas.93.20.11219

Bastos, A.M., Lundqvist, M., Waite, A.S., Kopell, N., and Miller, E.K. (2020). Layer and rhythm specificity for predictive routing. *Proc Natl Acad Sci U S A* 117, 31459-31469. doi: 10.1073/pnas.2014868117

Bastos, A.M., Usrey, W.M., Adams, R.A., Mangun, G.R., Fries, P., and Friston, K.J. (2012). Canonical microcircuits for predictive coding. *Neuron* 76, 695-711. doi: 10.1016/j.neuron.2012.10.038

Bastos, A.M., Vezoli, J., Bosman, C.A., Schoffelen, J.M., Oostenveld, R., Dowdall, J.R., et al. (2015). Visual areas exert feedforward and feedback influences through distinct frequency channels. *Neuron* 85, 390-401. doi: 10.1016/j.neuron.2014.12.018

Bourdillon, P., Ren, L., Halgren, M., Paulk, A.C., Salami, P., Ulbert, I., et al. (2024). Differential cortical layer engagement during seizure initiation and spread in humans. *Nat Commun* 15, 5153. doi: 10.1038/s41467-024-48746-8

Bowles, S., Hickman, J., Peng, X., Williamson, W.R., Huang, R., Washington, K., et al. (2022). Vagus nerve stimulation drives selective circuit modulation through cholinergic reinforcement. *Neuron* 110, 2867-2885 e2867. doi: 10.1016/j.neuron.2022.06.017

Brennan, K.C., and Pietrobon, D. (2018). A Systems Neuroscience Approach to Migraine. *Neuron* 97, 1004-1021. doi: 10.1016/j.neuron.2018.01.029

Broncel, A., Bocian, R., Kłos-Wojtczak, P., Kulbat-Warycha, K., and Konopacki, J. (2020). Vagal nerve stimulation as a promising tool in the improvement of cognitive disorders. *Brain Research Bulletin* 155, 37-47. doi: 10.1016/j.brainresbull.2019.11.011

Brougher, J., Sanchez, C.A., Aziz, U.S., Gove, K.F., and Thorn, C.A. (2021). Vagus Nerve Stimulation Induced Motor Map Plasticity Does Not Require Cortical Dopamine. *Frontiers in Neuroscience* 15. doi: 10.3389/fnins.2021.693140

Carandini, M., and Heeger, D.J. (2011). Normalization as a canonical neural computation. *Nat Rev Neurosci* 13, 51-62. doi: 10.1038/nrn3136

Carreno, F.R., and Frazer, A. (2017). Vagal Nerve Stimulation for Treatment-Resistant Depression. *Neurotherapeutics* 14, 716-727. doi: 10.1007/s13311-017-0537-8

Carroll, A.M., Pruitt, D.T., Riley, J.R., Danaphongse, T.T., Rennaker, R.L., Engineer, C.T., et al. (2024). Vagus nerve stimulation during training fails to improve learning in healthy rats. *Sci Rep* 14, 18955. doi: 10.1038/s41598-024-69666-z

Cavanagh, J.F., and Frank, M.J. (2014). Frontal theta as a mechanism for cognitive control. *Trends Cogn Sci* 18, 414-421. doi: 10.1016/j.tics.2014.04.012

Chao, Z.C., Takaura, K., Wang, L., Fujii, N., and Dehaene, S. (2018). Large-Scale Cortical Networks for Hierarchical Prediction and Prediction Error in the Primate Brain. *Neuron* 100, 1252-1266 e1253. doi: 10.1016/j.neuron.2018.10.004

Clark, K.B., Naritoku, D.K., Smith, D.C., Browning, R.A., and Jensen, R.A. (1999). Enhanced recognition memory following vagus nerve stimulation in human subjects. *Nat Neurosci* 2, 94-98. doi: 10.1038/4600

Clifford, H.J., Paranathala, M.P., Wang, Y., Thomas, R.H., da Silva Costa, T., Duncan, J.S., et al. (2024). Vagus nerve stimulation for epilepsy: A narrative review of factors predictive of response. *Epilepsia*. doi: 10.1111/epi.18153

Collins, L., Boddington, L., Steffan, P.J., and McCormick, D. (2021). Vagus nerve stimulation induces widespread cortical and behavioral activation. *Curr Biol* 31, 2088-2098 e2083. doi: 10.1016/j.cub.2021.02.049

Dawson, J., Liu, C.Y., Francisco, G.E., Cramer, S.C., Wolf, S.L., Dixit, A., et al. (2021). Vagus nerve stimulation paired with rehabilitation for upper limb motor function after ischaemic stroke (VNS-REHAB): a randomised, blinded, pivotal, device trial. *Lancet* 397, 1545-1553. doi: 10.1016/S0140-6736(21)00475-X

Dawson, J., Pierce, D., Dixit, A., Kimberley, T.J., Robertson, M., Tarver, B., et al. (2016). Safety, Feasibility, and Efficacy of Vagus Nerve Stimulation Paired With Upper-Limb Rehabilitation After Ischemic Stroke. *Stroke* 47, 143-150. doi: 10.1161/STROKEAHA.115.010477

De Ridder, D., Langguth, B., and Vanneste, S. (2021). "Chapter 20 - Vagus nerve stimulation for tinnitus: A review and perspective," in *Progress in Brain Research*, eds. B. Langguth, T. Kleinjung, D. De Ridder, W. Schlee, and S. VannesteElsevier), 451-467.

Devilbiss, D.M., and Waterhouse, B.D. (2000). Norepinephrine exhibits two distinct profiles of action on sensory cortical neuron responses to excitatory synaptic stimuli. *Synapse* 37, 273-282. doi: 10.1002/1098-2396(20000915)37:4<273::AID-SYN4>3.0.CO;2-#

Dhamoon, M.S., Moon, Y.P., Paik, M.C., Boden-Albala, B., Rundek, T., Sacco, R.L., et al. (2009). Long-

term functional recovery after first ischemic stroke: the Northern Manhattan Study. *Stroke* 40, 2805-2811. doi: 10.1161/STROKEAHA.109.549576

Edeline, J.M., Manunta, Y., and Hennevin, E. (2011). Induction of selective plasticity in the frequency tuning of auditory cortex and auditory thalamus neurons by locus coeruleus stimulation. *Hear Res* 274, 75-84. doi: 10.1016/j.heares.2010.08.005

Eggermont, J.J. (2015). The auditory cortex and tinnitus – a review of animal and human studies. *European Journal of Neuroscience* 41, 665-676. doi: 10.1111/ejn.12759

Engineer, C.T., Shetake, J.A., Engineer, N.D., Vrana, W.A., Wolf, J.T., and Kilgard, M.P. (2017). Temporal plasticity in auditory cortex improves neural discrimination of speech sounds. *Brain Stimul* 10, 543-552. doi: 10.1016/j.brs.2017.01.007

Engineer, N.D., Kimberley, T.J., Prudente, C.N., Dawson, J., Tarver, W.B., and Hays, S.A. (2019). Targeted Vagus Nerve Stimulation for Rehabilitation After Stroke. *Front Neurosci* 13, 280. doi: 10.3389/fnins.2019.00280

Engineer, N.D., Riley, J.R., Seale, J.D., Vrana, W.A., Shetake, J.A., Sudanagunta, S.P., et al. (2011). Reversing pathological neural activity using targeted plasticity. *Nature* 470, 101-104. doi: 10.1038/nature09656

Englot, D.J., Chang, E.F., and Auguste, K.I. (2011). Vagus nerve stimulation for epilepsy: a meta-analysis of efficacy and predictors of response. *Journal of Neurosurgery* 115, 1248-1255. doi: 10.3171/2011.7.Jns11977

Ernst, L.D., Steffan, P.J., Srikanth, P., Wiedrick, J., Spencer, D.C., Datta, P., et al. (2023). Electrocorticography Analysis in Patients With Dual Neurostimulators Supports Desynchronization as a Mechanism of Action for Acute Vagal Nerve Stimulator Stimulation. *J Clin Neurophysiol* 40, 37-44. doi: 10.1097/WNP.0000000000000847

Fang, J., Rong, P., Hong, Y., Fan, Y., Liu, J., Wang, H., et al. (2016). Transcutaneous Vagus Nerve Stimulation Modulates Default Mode Network in Major Depressive Disorder. *Biol Psychiatry* 79, 266-273. doi: 10.1016/j.biopsych.2015.03.025

Farrand, A., Jacquemet, V., Verner, R., Owens, M., and Beaumont, E. (2023). Vagus nerve stimulation parameters evoke differential neuronal responses in the locus coeruleus. *Physiol Rep* 11, e15633. doi: 10.14814/phy2.15633

Felleman, D.J., and Van Essen, D.C. (1991). Distributed hierarchical processing in the primate cerebral cortex. *Cereb Cortex* 1, 1-47. doi: 10.1093/cercor/1.1.1-a

Flor, H., Elbert, T., Knecht, S., Wienbruch, C., Pantev, C., Birbaumers, N., et al. (1995). Phantom-limb pain as a perceptual correlate of cortical reorganization following arm amputation. *Nature* 375, 482-

484. doi: 10.1038/375482a0

Fornai, F., Ruffoli, R., Giorgi, F.S., and Paparelli, A. (2011). The role of locus coeruleus in the antiepileptic activity induced by vagus nerve stimulation. *Eur J Neurosci* 33, 2169-2178. doi: 10.1111/j.1460-9568.2011.07707.x

Francisco, G.E., Engineer, N.D., Dawson, J., Kimberley, T.J., Cramer, S.C., Prudente, C.N., et al. (2023). Vagus Nerve Stimulation Paired With Upper-Limb Rehabilitation After Stroke: 2- and 3-Year Follow-up From the Pilot Study. *Arch Phys Med Rehabil* 104, 1180-1187. doi: 10.1016/j.apmr.2023.02.012

Frangos, E., and Komisaruk, B.R. (2017). Access to Vagal Projections via Cutaneous Electrical Stimulation of the Neck: fMRI Evidence in Healthy Humans. *Brain Stimul* 10, 19-27. doi: 10.1016/j.brs.2016.10.008

Fries, P. (2023). Rhythmic attentional scanning. *Neuron* 111, 954-970. doi: 10.1016/j.neuron.2023.02.015

Friston, K., and Kiebel, S. (2009). Predictive coding under the free-energy principle. *Philos Trans R Soc Lond B Biol Sci* 364, 1211-1221. doi: 10.1098/rstb.2008.0300

Froemke, R.C. (2015). Plasticity of cortical excitatory-inhibitory balance. *Annu Rev Neurosci* 38, 195-219. doi: 10.1146/annurev-neuro-071714-034002

Froemke, R.C., Carcea, I., Barker, A.J., Yuan, K., Seybold, B.A., Martins, A.R., et al. (2013). Long-term modification of cortical synapses improves sensory perception. *Nat Neurosci* 16, 79-88. doi: 10.1038/nn.3274

Froemke, R.C., Merzenich, M.M., and Schreiner, C.E. (2007). A synaptic memory trace for cortical receptive field plasticity. *Nature* 450, 425-429. doi: 10.1038/nature06289

Furmaga, H., Shah, A., and Frazer, A. (2011). Serotonergic and noradrenergic pathways are required for the anxiolytic-like and antidepressant-like behavioral effects of repeated vagal nerve stimulation in rats. *Biol Psychiatry* 70, 937-945. doi: 10.1016/j.biopsych.2011.07.020

George, M., Sonnen, A., Upton, A., Salinsky, M., Ristanovic, R., Bergen, D., et al. (1995). A randomized controlled trial of chronic vagus nerve stimulation for treatment of medically intractable seizures. *Neurology* 45, 224-230. doi: 10.1212/wnl.45.2.224

Ghacibeh, G.A., Shenker, J.I., Shenal, B., Uthman, B.M., and Heilman, K.M. (2006). Effect of vagus nerve stimulation on creativity and cognitive flexibility. *Epilepsy Behav* 8, 720-725. doi: 10.1016/j.yebeh.2006.03.008

Glennon, E., Valtcheva, S., Zhu, A., Wadghiri, Y.Z., Svirsky, M.A., and Froemke, R.C. (2023). Locus coeruleus activity improves cochlear implant performance. *Nature* 613, 317-323. doi: 10.1038/s41586-022-05554-8

Goadsby, P.J., Holland, P.R., Martins-Oliveira, M., Hoffmann, J., Schankin, C., and Akerman, S. (2017). Pathophysiology of Migraine: A Disorder of Sensory Processing. *Physiol Rev* 97, 553-622. doi: 10.1152/physrev.00034.2015

Goard, M., and Dan, Y. (2009). Basal forebrain activation enhances cortical coding of natural scenes. *Nat Neurosci* 12, 1444-1449. doi: 10.1038/nn.2402

Grimm, C., Duss, S.N., Privitera, M., Munn, B.R., Karalis, N., Frassle, S., et al. (2024). Tonic and burst-like locus coeruleus stimulation distinctly shift network activity across the cortical hierarchy. *Nat Neurosci* 27, 2167-2177. doi: 10.1038/s41593-024-01755-8

Grimonprez, A., Raedt, R., Baeken, C., Boon, P., and Vonck, K. (2015). The antidepressant mechanism of action of vagus nerve stimulation: Evidence from preclinical studies. *Neurosci Biobehav Rev* 56, 26-34. doi: 10.1016/j.neubiorev.2015.06.019

Halvagal, M.S., and Zenke, F. (2023). The combination of Hebbian and predictive plasticity learns invariant object representations in deep sensory networks. *Nat Neurosci* 26, 1906-1915. doi: 10.1038/s41593-023-01460-y

Hamilton, J.P., Farmer, M., Fogelman, P., and Gotlib, I.H. (2015). Depressive Rumination, the Default-Mode Network, and the Dark Matter of Clinical Neuroscience. *Biol Psychiatry* 78, 224-230. doi: 10.1016/j.biopsych.2015.02.020

Hamilton, J.P., Furman, D.J., Chang, C., Thomason, M.E., Dennis, E., and Gotlib, I.H. (2011). Default-mode and task-positive network activity in major depressive disorder: implications for adaptive and maladaptive rumination. *Biol Psychiatry* 70, 327-333. doi: 10.1016/j.biopsych.2011.02.003

Handforth, A., DeGiorgio, C.M., Schachter, S.C., Uthman, B.M., Naritoku, D.K., Tecoma, E.S., et al. (1998). Vagus nerve stimulation therapy for partial-onset seizures: a randomized active-control trial. *Neurology* 51, 48-55. doi: 10.1212/wnl.51.1.48

Hays, S.A., Rennaker, R.L., and Kilgard, M.P. (2023). How to fail with paired VNS therapy. *Brain Stimulation* 16, 1252-1258. doi: 10.1016/j.brs.2023.08.009

Heilbron, M., and Chait, M. (2018). Great Expectations: Is there Evidence for Predictive Coding in Auditory Cortex? *Neuroscience* 389, 54-73. doi: 10.1016/j.neuroscience.2017.07.061

Helmstaedter, C., Hoppe, C., and Elger, C.E. (2001). Memory alterations during acute high-intensity vagus nerve stimulation. *Epilepsy Res* 47, 37-42. doi: 10.1016/s0920-1211(01)00291-1

Henry, T.R., Votaw, J.R., Pennell, P.B., Epstein, C.M., Bakay, R.A., Faber, T.L., et al. (1999). Acute blood flow changes and efficacy of vagus nerve stimulation in partial epilepsy. *Neurology* 52, 1166-1173. doi: 10.1212/wnl.52.6.1166

Hulsey, D.R., Hays, S.A., Khodaparast, N., Ruiz, A., Das, P., Rennaker, R.L., 2nd, et al. (2016).

Reorganization of Motor Cortex by Vagus Nerve Stimulation Requires Cholinergic Innervation. *Brain Stimul* 9, 174-181. doi: 10.1016/j.brs.2015.12.007

Hulsey, D.R., Shedd, C.M., Sarker, S.F., Kilgard, M.P., and Hays, S.A. (2019). Norepinephrine and serotonin are required for vagus nerve stimulation directed cortical plasticity. *Exp Neurol* 320, 112975. doi: 10.1016/j.expneurol.2019.112975

Ibrahim, G.M., Sharma, P., Hyslop, A., Guillen, M.R., Morgan, B.R., Wong, S., et al. (2017). Presurgical thalamocortical connectivity is associated with response to vagus nerve stimulation in children with intractable epilepsy. *Neuroimage Clin* 16, 634-642. doi: 10.1016/j.nicl.2017.09.015

Ishizu, K., Shiramatsu, T.I., Hitsuyu, R., Oizumi, M., Tsuchiya, N., and Takahashi, H. (2021). Information flow in the rat thalamo-cortical system: spontaneous vs. stimulus-evoked activities. *Sci Rep* 11, 19252. doi: 10.1038/s41598-021-98660-y

Kaiser, R.H., Andrews-Hanna, J.R., Wager, T.D., and Pizzagalli, D.A. (2015). Large-Scale Network Dysfunction in Major Depressive Disorder: A Meta-analysis of Resting-State Functional Connectivity. *JAMA Psychiatry* 72, 603-611. doi: 10.1001/jamapsychiatry.2015.0071

Kawai, K., Tanaka, T., Baba, H., Bunker, M., Ikeda, A., Inoue, Y., et al. (2017). Outcome of vagus nerve stimulation for drug-resistant epilepsy: the first three years of a prospective Japanese registry. *Epileptic Disord* 19, 327-338. doi: 10.1684/epd.2017.0929

Kawakami, S., Okada, N., Satomura, Y., Shoji, E., Mori, S., Kiyota, M., et al. (2024). Frontal pole-precuneus connectivity is associated with a discrepancy between self-rated and observer-rated depression severity in mood disorders: a resting-state functional magnetic resonance imaging study. *Cerebral Cortex* 34. doi: ARTN bhae284

10.1093/cercor/bhae284

Khodaparast, N., Hays, S.A., Sloan, A.M., Fayyaz, T., Hulsey, D.R., Rennaker, R.L., 2nd, et al. (2014). Vagus nerve stimulation delivered during motor rehabilitation improves recovery in a rat model of stroke. *Neurorehabil Neural Repair* 28, 698-706. doi: 10.1177/1545968314521006

Kilgard, M.P., and Merzenich, M.M. (1998a). Cortical map reorganization enabled by nucleus basalis activity. *Science* 279, 1714-1718. doi: 10.1126/science.279.5357.1714

Kilgard, M.P., and Merzenich, M.M. (1998b). Plasticity of temporal information processing in the primary auditory cortex. *Nat Neurosci* 1, 727-731. doi: 10.1038/3729

Kimberley, T.J., Pierce, D., Prudente, C.N., Francisco, G.E., Yozbatiran, N., Smith, P., et al. (2018). Vagus Nerve Stimulation Paired With Upper Limb Rehabilitation After Chronic Stroke. *Stroke* 49, 2789-2792. doi: 10.1161/STROKEAHA.118.022279

Klaming, R., Simmons, A.N., Spadoni, A.D., and Lerman, I. (2022). Effects of Noninvasive Cervical Vagal

Nerve Stimulation on Cognitive Performance But Not Brain Activation in Healthy Adults. *Neuromodulation* 25, 424-432. doi: 10.1111/ner.13313

Kong, Y., Zhao, K., Zeng, D., Lu, F., Li, X., Wu, Y., et al. (2024). Effects of vagus nerve stimulation on cognitive function in patients with epilepsy: a systematic review and meta-analysis. *Front Neurol* 15, 1332882. doi: 10.3389/fneur.2024.1332882

Krahl, S.E., Clark, K.B., Smith, D.C., and Browning, R.A. (1998). Locus coeruleus lesions suppress the seizure-attenuating effects of vagus nerve stimulation. *Epilepsia* 39, 709-714. doi: 10.1111/j.1528-1157.1998.tb01155.x

Kumagai, S., Shiramatsu, T.I., Matsumura, A., Ishishita, Y., Ibayashi, K., Onuki, Y., et al. (2023). Frequency-specific modulation of oscillatory activity in the rat auditory cortex by vagus nerve stimulation. *Brain Stimul* 16, 1476-1485. doi: 10.1016/j.brs.2023.09.019

Kwakkel, G., and Dobkin, B.H. (2021). Vagus Nerve Stimulation for Upper Limb Function. *Stroke* 52, 3407-3409. doi: 10.1161/strokeaha.121.035648

Laszlovszky, T., Schlingloff, D., Hegedüs, P., Freund, T.F., Gulyás, A., Kepecs, A., et al. (2020). Distinct synchronization, cortical coupling and behavioral function of two basal forebrain cholinergic neuron types. *Nature Neuroscience* 23, 992-1003. doi: 10.1038/s41593-020-0648-0

Lewine, J.D., Paulson, K., Bangera, N., and Simon, B.J. (2019). Exploration of the Impact of Brief Noninvasive Vagal Nerve Stimulation on EEG and Event-Related Potentials. *Neuromodulation* 22, 564-572. doi: 10.1111/ner.12864

Liu, Y., Jia, W., Strosberg, A.D., and Cynader, M. (1993). Development and regulation of β adrenergic receptors in kitten visual cortex: An immunocytochemical and autoradiographic study. *Brain Research* 632, 274-286. doi: 10.1016/0006-8993(93)91162-l

Lloyd, B., Wurm, F., de Kleijn, R., and Nieuwenhuis, S. (2023). Short-term transcutaneous vagus nerve stimulation increases pupil size but does not affect EEG alpha power: A replication of Sharon et al. (2021, Journal of Neuroscience). *Brain Stimulation* 16, 1001-1008. doi: 10.1016/j.brs.2023.06.010

Lohani, S., Moberly, A.H., Benisty, H., Landa, B., Jing, M., Li, Y., et al. (2022). Spatiotemporally heterogeneous coordination of cholinergic and neocortical activity. *Nat Neurosci* 25, 1706-1713. doi: 10.1038/s41593-022-01202-6

Lu, F.M., Cui, Q., Huang, X.J., Li, L.Y., Duan, X.J., Chen, H., et al. (2020). Anomalous intrinsic connectivity within and between visual and auditory networks in major depressive disorder. *Prog Neuro-Psychoph* 100, 109889. doi: ARTN 109889

10.1016/j.pnpbp.2020.109889

Mahadi, K.M., Lall, V.K., Deuchars, S.A., and Deuchars, J. (2019). Cardiovascular autonomic effects of transcutaneous auricular nerve stimulation via the tragus in the rat involve spinal cervical sensory afferent pathways. *Brain Stimul* 12, 1151-1158. doi: 10.1016/j.brs.2019.05.002

Makino, H., and Komiyama, T. (2015). Learning enhances the relative impact of top-down processing in the visual cortex. *Nat Neurosci* 18, 1116-1122. doi: 10.1038/nn.4061

Malakouti, N., Serruya, M.D., Cramer, S.C., Kimberley, T.J., and Rosenwasser, R.H. (2024). Making Sense of Vagus Nerve Stimulation for Stroke. *Stroke* 55, 519-522. doi: 10.1161/STROKEAHA.123.044576

Markov, N.T., Ercsey-Ravasz, M., Van Essen, D.C., Knoblauch, K., Toroczkai, Z., and Kennedy, H. (2013). Cortical High-Density Counterstream Architectures. *Science* 342, 1238406. doi: doi:10.1126/science.1238406

Markov, N.T., Vezoli, J., Chameau, P., Falchier, A., Quilodran, R., Huissoud, C., et al. (2014). Anatomy of hierarchy: feedforward and feedback pathways in macaque visual cortex. *J Comp Neurol* 522, 225-259. doi: 10.1002/cne.23458

Martelletti, P., Barbanti, P., Grazzi, L., Pierangeli, G., Rainero, I., Geppetti, P., et al. (2018). Consistent effects of non-invasive vagus nerve stimulation (nVNS) for the acute treatment of migraine: additional findings from the randomized, sham-controlled, double-blind PRESTO trial. *The Journal of Headache and Pain* 19. doi: 10.1186/s10194-018-0929-0

Martin, K.A., Papadoyannis, E.S., Schiavo, J.K., Fadaei, S.S., Issa, H.A., Song, S.C., et al. (2024). Vagus nerve stimulation recruits the central cholinergic system to enhance perceptual learning. *Nat Neurosci* 27, 2152-2166. doi: 10.1038/s41593-024-01767-4

Martins, A.R., and Froemke, R.C. (2015). Coordinated forms of noradrenergic plasticity in the locus coeruleus and primary auditory cortex. *Nat Neurosci* 18, 1483-1492. doi: 10.1038/nn.4090

Marx, W., Penninx, B.W.J.H., Solmi, M., Furukawa, T.A., Firth, J., Carvalho, A.F., et al. (2023). Major depressive disorder. *Nature Reviews Disease Primers* 9. doi: 10.1038/s41572-023-00454-1

McGlone, J., Valdivia, I., Penner, M., Williams, J., Sadler, R.M., and Clarke, D.B. (2008). Quality of life and memory after vagus nerve stimulator implantation for epilepsy. *Can J Neurol Sci* 35, 287-296. doi: 10.1017/s0317167100008854

McIntyre, R.S., Alsuwaidan, M., Baune, B.T., Berk, M., Demyttenaere, K., Goldberg, J.F., et al. (2023). Treatment-resistant depression: definition, prevalence, detection, management, and investigational interventions. *World Psychiatry* 22, 394-412. doi: 10.1002/wps.21120

McLin, D.E., 3rd, Miasnikov, A.A., and Weinberger, N.M. (2002). Induction of behavioral associative memory by stimulation of the nucleus basalis. *Proc Natl Acad Sci U S A* 99, 4002-4007. doi:

10.1073/pnas.062057099

Mertens, A., Gadeyne, S., Lescrauwaet, E., Carrette, E., Meurs, A., De Herdt, V., et al. (2022). The potential of invasive and non-invasive vagus nerve stimulation to improve verbal memory performance in epilepsy patients. *Sci Rep* 12, 1984. doi: 10.1038/s41598-022-05842-3

Mertens, A., Naert, L., Miatton, M., Poppa, T., Carrette, E., Gadeyne, S., et al. (2020). Transcutaneous Vagus Nerve Stimulation Does Not Affect Verbal Memory Performance in Healthy Volunteers. *Frontiers in Psychology* 11. doi: 10.3389/fpsyg.2020.00551

Meyers, E.C., Kasliwal, N., Solorzano, B.R., Lai, E., Bendale, G., Berry, A., et al. (2019). Enhancing plasticity in central networks improves motor and sensory recovery after nerve damage. *Nat Commun* 10, 5782. doi: 10.1038/s41467-019-13695-0

Meyers, E.C., Solorzano, B.R., James, J., Ganzer, P.D., Lai, E.S., Rennaker, R.L., 2nd, et al. (2018). Vagus Nerve Stimulation Enhances Stable Plasticity and Generalization of Stroke Recovery. *Stroke* 49, 710-717. doi: 10.1161/STROKEAHA.117.019202

Miller, E.K., Lundqvist, M., and Bastos, A.M. (2018). Working Memory 2.0. *Neuron* 100, 463-475. doi: 10.1016/j.neuron.2018.09.023

Mithani, K., Mikhail, M., Morgan, B.R., Wong, S., Weil, A.G., Deschenes, S., et al. (2019). Connectomic Profiling Identifies Responders to Vagus Nerve Stimulation. *Ann Neurol* 86, 743-753. doi: 10.1002/ana.25574

Morrison, R.A., Hulsey, D.R., Adcock, K.S., Rennaker, R.L., 2nd, Kilgard, M.P., and Hays, S.A. (2019). Vagus nerve stimulation intensity influences motor cortex plasticity. *Brain Stimul* 12, 256-262. doi: 10.1016/j.brs.2018.10.017

Mühlnickel, W., Elbert, T., Taub, E., and Flor, H. (1998). Reorganization of auditory cortex in tinnitus. *Proceedings of the National Academy of Sciences* 95, 10340-10343. doi: 10.1073/pnas.95.17.10340

Munn, B.R., Muller, E.J., Wainstein, G., and Shine, J.M. (2021). The ascending arousal system shapes neural dynamics to mediate awareness of cognitive states. *Nat Commun* 12, 6016. doi: 10.1038/s41467-021-26268-x

Nemeroff, C.B., Mayberg, H.S., Krahl, S.E., McNamara, J., Frazer, A., Henry, T.R., et al. (2006). VNS therapy in treatment-resistant depression: clinical evidence and putative neurobiological mechanisms. *Neuropsychopharmacology* 31, 1345-1355. doi: 10.1038/sj.npp.1301082

Owens, M.M., Jacquemet, V., Napadow, V., Lewis, N., and Beaumont, E. (2024). Brainstem neuronal responses to transcutaneous auricular and cervical vagus nerve stimulation in rats. *J Physiol* 602, 4027-4052. doi: 10.1113/JP286680

Palomero-Gallagher, N., and Zilles, K. (2019). Cortical layers: Cyto-, myelo-, receptor- and synaptic architecture in human cortical areas. *Neuroimage* 197, 716-741. doi: 10.1016/j.neuroimage.2017.08.035

Porter, B.A., Khodaparast, N., Fayyaz, T., Cheung, R.J., Ahmed, S.S., Vrana, W.A., et al. (2012). Repeatedly pairing vagus nerve stimulation with a movement reorganizes primary motor cortex. *Cereb Cortex* 22, 2365-2374. doi: 10.1093/cercor/bhr316

Puledda, F., Silva, E.M., Suwanlaong, K., and Goadsby, P.J. (2023). Migraine: from pathophysiology to treatment. *J Neurol* 270, 3654-3666. doi: 10.1007/s00415-023-11706-1

Raedt, R., Clinckers, R., Mollet, L., Vonck, K., El Tahry, R., Wyckhuys, T., et al. (2011). Increased hippocampal noradrenaline is a biomarker for efficacy of vagus nerve stimulation in a limbic seizure model. *J Neurochem* 117, 461-469. doi: 10.1111/j.1471-4159.2011.07214.x

Ramos, B.P., Colgan, L., Nou, E., Ovadia, S., Wilson, S.R., and Arnsten, A.F. (2005). The beta-1 adrenergic antagonist, betaxolol, improves working memory performance in rats and monkeys. *Biol Psychiatry* 58, 894-900. doi: 10.1016/j.biopsych.2005.05.022

Reed, A., Riley, J., Carraway, R., Carrasco, A., Perez, C., Jakkamsetti, V., et al. (2011). Cortical Map Plasticity Improves Learning but Is Not Necessary for Improved Performance. *Neuron* 70, 121-131. doi: 10.1016/j.neuron.2011.02.038

Ridgewell, C., Heaton, K.J., Hildebrandt, A., Couse, J., Leeder, T., and Neumeier, W.H. (2021). The effects of transcutaneous auricular vagal nerve stimulation on cognition in healthy individuals: A meta-analysis. *Neuropsychology* 35, 352-365. doi: 10.1037/neu0000735

Rodenkirch, C., and Wang, Q. (2020). Rapid and transient enhancement of thalamic information transmission induced by vagus nerve stimulation. *J Neural Eng* 17, 026027. doi: 10.1088/1741-2552/ab6b84

Rush, A.J., Sackeim, H.A., Marangell, L.B., George, M.S., Brannan, S.K., Davis, S.M., et al. (2005). Effects of 12 months of vagus nerve stimulation in treatment-resistant depression: a naturalistic study. *Biol Psychiatry* 58, 355-363. doi: 10.1016/j.biopsych.2005.05.024

Salgado, H., Köhr, G., and Treviño, M. (2012). Noradrenergic 'tone' determines dichotomous control of cortical spike-timing-dependent plasticity. *Sci Rep* 2, 417. doi: 10.1038/srep00417

Salinsky, M.C., and Burchiel, K.J. (1993). Vagus nerve stimulation has no effect on awake EEG rhythms in humans. *Epilepsia* 34, 299-304. doi: 10.1111/j.1528-1157.1993.tb02415.x

Saponati, M., and Vinck, M. (2023). Sequence anticipation and spike-timing-dependent plasticity emerge from a predictive learning rule. *Nat Commun* 14, 4985. doi: 10.1038/s41467-023-40651-w

Sarnthein, J., Petsche, H., Rappelsberger, P., Shaw, G.L., and von Stein, A. (1998). Synchronization

between prefrontal and posterior association cortex during human working memory. *Proceedings of the National Academy of Sciences* 95, 7092-7096. doi: 10.1073/pnas.95.12.7092

Schlaepfer, T.E., Frick, C., Zobel, A., Maier, W., Heuser, I., Bajbouj, M., et al. (2008). Vagus nerve stimulation for depression: efficacy and safety in a European study. *Psychological Medicine* 38, 651-661. doi: 10.1017/s0033291707001924

Schmitz, T.W., and Duncan, J. (2018). Normalization and the Cholinergic Microcircuit: A Unified Basis for Attention. *Trends Cogn Sci* 22, 422-437. doi: 10.1016/j.tics.2018.02.011

Schuerman, W.L., Nourski, K.V., Rhone, A.E., Howard, M.A., Chang, E.F., and Leonard, M.K. (2021). Human intracranial recordings reveal distinct cortical activity patterns during invasive and non-invasive vagus nerve stimulation. *Sci Rep* 11, 22780. doi: 10.1038/s41598-021-02307-x

Sharon, O., Fahoum, F., and Nir, Y. (2021). Transcutaneous Vagus Nerve Stimulation in Humans Induces Pupil Dilation and Attenuates Alpha Oscillations. *J Neurosci* 41, 320-330. doi: 10.1523/JNEUROSCI.1361-20.2020

Shine, J.M. (2019). Neuromodulatory Influences on Integration and Segregation in the Brain. *Trends Cogn Sci* 23, 572-583. doi: 10.1016/j.tics.2019.04.002

Shine, J.M., Muller, E.J., Munn, B., Cabral, J., Moran, R.J., and Breakspear, M. (2021). Computational models link cellular mechanisms of neuromodulation to large-scale neural dynamics. *Nat Neurosci* 24, 765-776. doi: 10.1038/s41593-021-00824-6

Shipp, S. (2007). Structure and function of the cerebral cortex. *Curr Biol* 17, R443-449. doi: 10.1016/j.cub.2007.03.044

Takahashi, H., Funamizu, A., Mitsumori, Y., Kose, H., and Kanzaki, R. (2010). Progressive plasticity of auditory cortex during appetitive operant conditioning. *Biosystems* 101, 37-41. doi: 10.1016/j.biosystems.2010.04.003

Takahashi, H., Shiramatsu, T.I., Hitsuyu, R., Ibayashi, K., and Kawai, K. (2020). Vagus nerve stimulation (VNS)-induced layer-specific modulation of evoked responses in the sensory cortex of rats. *Sci Rep* 10, 8932. doi: 10.1038/s41598-020-65745-z

Takahashi, H., Yokota, R., Funamizu, A., Kose, H., and Kanzaki, R. (2011). Learning-stage-dependent, field-specific, map plasticity in the rat auditory cortex during appetitive operant conditioning. *Neuroscience* 199, 243-258. doi: 10.1016/j.neuroscience.2011.09.046

Takahashi, H., Yokota, R., and Kanzaki, R. (2013). Response variance in functional maps: neural darwinism revisited. *PLoS One* 8, e68705. doi: 10.1371/journal.pone.0068705

Tassorelli, C., Grazzi, L., de Tommaso, M., Pierangeli, G., Martelletti, P., Rainero, I., et al. (2018). Noninvasive vagus nerve stimulation as acute therapy for migraine: The randomized PRESTO

study. *Neurology* 91, e364-e373. doi: 10.1212/WNL.0000000000005857

Thakkar, V.J., Engelhart, A.S., Khodaparast, N., Abadzi, H., and Centanni, T.M. (2020). Transcutaneous auricular vagus nerve stimulation enhances learning of novel letter-sound relationships in adults. *Brain Stimul* 13, 1813-1820. doi: 10.1016/j.brs.2020.10.012

Tseng, C.T., Gaulding, S.J., Dancel, C.L.E., and Thorn, C.A. (2021). Local activation of alpha2 adrenergic receptors is required for vagus nerve stimulation induced motor cortical plasticity. *Sci Rep* 11, 21645. doi: 10.1038/s41598-021-00976-2

Vezoli, J., Vinck, M., Bosman, C.A., Bastos, A.M., Lewis, C.M., Kennedy, H., et al. (2021). Brain rhythms define distinct interaction networks with differential dependence on anatomy. *Neuron* 109, 3862-3878 e3865. doi: 10.1016/j.neuron.2021.09.052

Vinck, M., Batista-Brito, R., Knoblich, U., and Cardin, J.A. (2015). Arousal and locomotion make distinct contributions to cortical activity patterns and visual encoding. *Neuron* 86, 740-754. doi: 10.1016/j.neuron.2015.03.028

Vonck, K., Raedt, R., Naulaerts, J., De Vogelaere, F., Thiery, E., Van Roost, D., et al. (2014). Vagus nerve stimulation…25 years later! What do we know about the effects on cognition? *Neuroscience & Biobehavioral Reviews* 45, 63-71. doi: 10.1016/j.neubiorev.2014.05.005

Wake, N., Shiramatsu, T.I., and Takahashi, H. (2024). Map plasticity following noise exposure in auditory cortex of rats: implications for disentangling neural correlates of tinnitus and hyperacusis. *Front Neurosci* 18, 1385942. doi: 10.3389/fnins.2024.1385942

Wang, L., Zhang, J., Guo, C., He, J., Zhang, S., Wang, Y., et al. (2022). The efficacy and safety of transcutaneous auricular vagus nerve stimulation in patients with mild cognitive impairment: A double blinded randomized clinical trial. *Brain Stimul* 15, 1405-1414. doi: 10.1016/j.brs.2022.09.003

Weber, I., Niehaus, H., Krause, K., Molitor, L., Peper, M., Schmidt, L., et al. (2021). Trust your gut: vagal nerve stimulation in humans improves reinforcement learning. *Brain Commun* 3, fcab039. doi: 10.1093/braincomms/fcab039

Weinberger, N.M. (2003). The nucleus basalis and memory codes: auditory cortical plasticity and the induction of specific, associative behavioral memory. *Neurobiol Learn Mem* 80, 268-284. doi: 10.1016/s1074-7427(03)00072-8

Weinberger, N.M. (2004). Specific long-term memory traces in primary auditory cortex. *Nat Rev Neurosci* 5, 279-290. doi: 10.1038/nrn1366

Weinberger, N.M. (2007). Associative representational plasticity in the auditory cortex: a synthesis of two disciplines. *Learn Mem* 14, 1-16. doi: 10.1101/lm.421807

Workewych, A.M., Arski, O.N., Mithani, K., and Ibrahim, G.M. (2020). Biomarkers of seizure response to vagus nerve stimulation: A scoping review. *Epilepsia* 61, 2069-2085. doi: 10.1111/epi.16661

Yokoyama, R., Akiyama, Y., Enatsu, R., Suzuki, H., Suzuki, Y., Kanno, A., et al. (2020). The Immediate Effects of Vagus Nerve Stimulation in Intractable Epilepsy: An Intra-operative Electrocorticographic Analysis. *Neurol Med Chir (Tokyo)* 60, 244-251. doi: 10.2176/nmc.oa.2019-0221

Zhang, Y., Liu, J., Li, H., Yan, Z., Liu, X., Cao, J., et al. (2019). Transcutaneous auricular vagus nerve stimulation at 1 Hz modulates locus coeruleus activity and resting state functional connectivity in patients with migraine: An fMRI study. *Neuroimage Clin* 24, 101971. doi: 10.1016/j.nicl.2019.101971

Zuo, Y., Smith, D.C., and Jensen, R.A. (2007). Vagus nerve stimulation potentiates hippocampal LTP in freely-moving rats. *Physiol Behav* 90, 583-589. doi: 10.1016/j.physbeh.2006.11.009

**Figure legend**

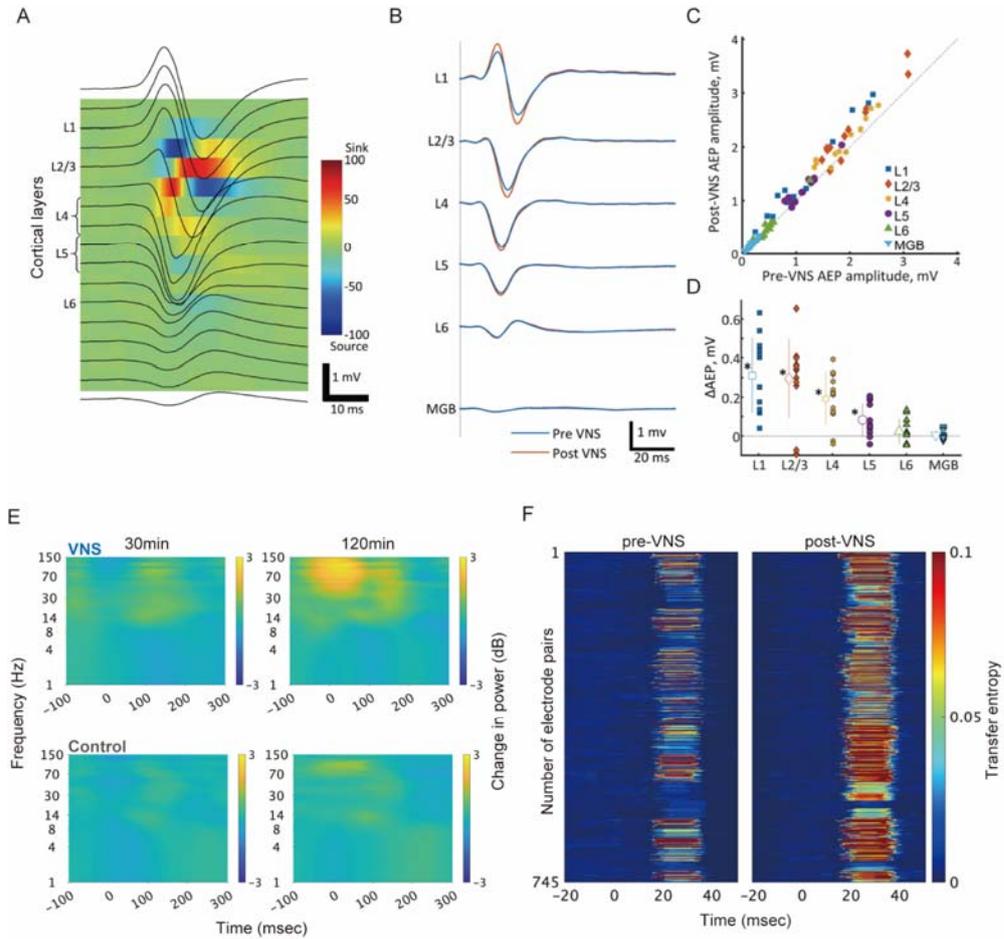

**FIGURE 1** VNS differentially modulates cortical processing across layers and pathways in the auditory system. (A) Current source density (CSD) analysis for layer identification in the rat auditory cortex. Representative laminar recordings showing auditory evoked potentials (AEPs, black traces) superimposed on the color-coded CSD map. The characteristic pattern of current sinks (red) and sources (blue) enabled precise identification of cortical layers L1-L6. Scale bar: 1mV, 10ms. (B) VNS-induced modulation of click-evoked responses across cortical layers. Grand-averaged AEP waveforms recorded simultaneously from different cortical layers before (blue) and after (red) VNS. Layer-specific enhancement was observed in response amplitudes. (C) Quantitative comparison of AEP amplitudes between pre- and post-VNS conditions. Scatter plot of AEP amplitudes showing individual recording sites across different layers, with each point representing pre- vs post-VNS amplitudes. Points above the diagonal line indicate VNS-induced enhancement. (D) Layer-specific profile of VNS effects. The magnitude of VNS-induced changes in AEP amplitude (ΔAEP) showed layer-dependent differences. Comparable enhancement was observed in superficial layers, whereas this effect diminished in deeper layers. Data points represent individual recordings; bars indicate mean ± SD.

Asterisks denote statistical significance (*$p$ < 0.05). (E) Time-frequency analysis of VNS-induced modulation of oscillatory activity in response to click sounds. Spectrograms comparing VNS (top) and control (bottom) groups showing the temporal evolution of frequency-specific power in the rat auditory core region at early (30 min) and late (120 min) timepoints during click presentation experiments. The color scale (right) indicates power changes in decibels relative to baseline period (-500 to -200 msec). The VNS group demonstrated progressive enhancement of high-frequency oscillations (gamma: 30-150 Hz) accompanied by attenuation of low-frequency activity (theta: 4-8 Hz). In contrast, the control group showed minimal changes in oscillatory patterns over time. Data were averaged across recording sites within the functionally defined auditory core region. (F) Information flow analysis reveals VNS-induced modulation of feedforward pathways in the rat auditory cortex. Transfer entropy (TE) analysis was performed on multi-unit recordings from functionally connected electrode pairs in the core and belt regions during click-evoked responses. Each row represents individual electrode pairs from a single rat. Left: Color-coded normalized TE values for feedforward pathways through layer 4 before VNS. Right: Corresponding TE values after VNS, demonstrating enhanced information flow. TE analysis was performed following the methods described in (Ishizu et al., 2021). CSD, current source density; AEP, auditory evoked potential; VNS, vagus nerve stimulation; MGB, medial geniculate body in the ventral division of thalamus; L, layer; TE, transfer entropy.

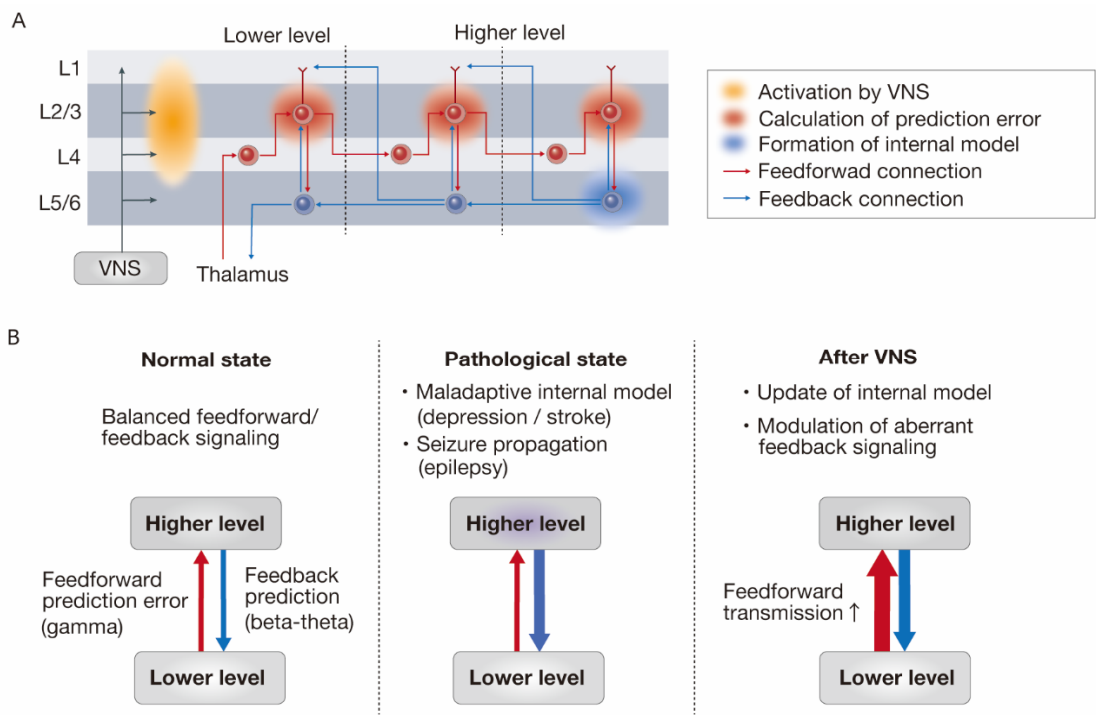

**FIGURE 2** VNS-induced modulation of neural information processing through feedforward-feedback interactions. (A) Cortical laminar organization of feedforward and feedback signals in prediction error processing. Prediction errors are computed in layer 2/3 (supragranular layers) through the integration of multiple inputs: feedback predictions from higher cortical areas arriving via layer 1, local predictions from layer 5/6 (infragranular layers), and feedforward sensory inputs from lower cortical areas. This hierarchical process allows layer 2/3 neurons to compare incoming sensory signals against predictions, enabling the computation of prediction errors through complex multi-layered processing. These computed prediction errors are then propagated in a feedforward manner, thereby continuously updating internal models in higher cortical areas, enabling adaptive predictions about incoming sensory information. VNS may predominantly enhance the activation of superficial layers compared to deep layers, potentially facilitating the feedforward transmission of prediction errors throughout the cortical hierarchy. (B) Schematic illustration of hierarchical information processing in normal, pathological, and VNS-treated states. In the normal state, there is a balanced interaction between feedforward prediction error (gamma) and feedback prediction (beta-theta) signaling. The pathological state is characterized by maladaptive internal model (depression / stroke) and seizure propagation (epilepsy). VNS therapy enhances feedforward transmission, which leads to updating of the internal model and modulation of aberrant feedback signaling. Red and blue arrows indicate feedforward and feedback signaling, respectively. VNS, vagus nerve stimulation.